\documentclass[final,3p,times,twocolumn]{elsarticle}
\usepackage{algorithm}
\usepackage{algpseudocode}
\usepackage{lineno}
\modulolinenumbers[5]

\journal{Structural and Multidisciplinary Optimization}
\usepackage[table,xcdraw]{xcolor}
\usepackage{graphicx}
\usepackage{subcaption}
\usepackage{dcolumn}
\usepackage{bm}
\usepackage{multirow} 
\usepackage{color}
\usepackage{mathtools}
\usepackage{multirow}
\usepackage{caption}
\usepackage{amsmath} 
\usepackage{chemformula} 
\usepackage{csquotes} 
\biboptions{numbers,sort&compress}
\usepackage{subfig}
\usepackage{placeins}
\usepackage{cancel}
\usepackage[normalem]{ulem}
\usepackage{siunitx}
\usepackage{bm}
\usepackage[capitalise]{cleveref}
\usepackage{xr}
\usepackage{comment}
\usepackage{csvsimple}
\usepackage{breqn}
\usepackage{amsmath}
\usepackage{array} 
\usepackage{capt-of}
\newcolumntype{P}[1]{>{\centering\arraybackslash}p{#1}}
\newcommand{\eref}[1]{Eq.~(\ref{#1})}

\newcommand{\sref}[1]{Section~\ref{#1}}

\makeatletter
\newcommand*{\addFileDependency}[1]{
  \typeout{(#1)}
  \@addtofilelist{#1}
  \IfFileExists{#1}{}{\typeout{No file #1.}}
}
\makeatother

\begin{document}

\begin{frontmatter}
\title{Physics-Informed Transformer for Real-Time High-Fidelity Topology Optimization}

\author[A]{Aaron Lutheran}
\author[B]{Srijan Das}
\author[A,C]{Alireza Tabarraei\corref{mycorrespondingauthor}}
\cortext[mycorrespondingauthor]{Corresponding author}
\ead{atabarra@charlotte.edu}

\address[A]{Department of Mechanical Engineering and Engineering Science, The University of North Carolina at Charlotte, Charlotte, NC 28223, USA}
\address[B]{Department of Computer Science, The University of North Carolina at Charlotte, Charlotte, NC 28223, USA}
\address[C]{School of Data Science, The University of North Carolina at Charlotte, Charlotte, NC 28223, USA}

\begin{abstract}
Topology optimization is used for the design of high-performance structures but remains fundamentally limited by its iterative nature, requiring repeated finite element analyses that prevent real-time deployment and large-scale design exploration. In this work, we introduce a physics-informed transformer architecture that directly learns a non-iterative mapping from boundary conditions, loading configurations, and derived physical fields to optimized structural topologies. By leveraging global self-attention, the proposed model captures long-range mechanical interactions that govern structural response, overcoming the locality limitations of convolutional architectures. A conditioning-token mechanism embeds global problem parameters, while spatially distributed stress and strain energy fields are encoded as patch tokens within a Vision Transformer framework. To ensure physical realism and manufacturability, we incorporate auxiliary loss functions that enforce volume constraints, load adherence, and structural connectivity through a differentiable formulation. The framework is further extended to dynamic loading scenarios using frequency-domain encoding and transfer learning, enabling efficient generalization from static to time-dependent problems. Comprehensive benchmarking demonstrates that the proposed model achieves fidelity beyond that of diffusion models, while requiring only a single forward pass, thereby eliminating iterative inference entirely. This establishes topology optimization as a real-time operator-learning problem, enabling high-fidelity structural design with significant reductions in computational cost.
\end{abstract}

\end{frontmatter}

\section{Introduction}\label{introduction}

Topology optimization (TO) has become a cornerstone of modern computational design, enabling the systematic discovery of high-performance structures beyond the limits of conventional engineering intuition \cite{bendsoe2003topology, sigmund_topology_2013, deaton2014survey}. By optimizing the spatial distribution of material within a prescribed domain, TO provides a principled framework for achieving extreme performance objectives such as minimum compliance, maximum stiffness-to-weight ratio, and multifunctional efficiency. Its impact spans aerospace, automotive, civil engineering, biomedical devices, and additive manufacturing, where geometric freedom enables unprecedented structural performance \cite{Zhu2021Review, Meng2019Roadmap}. Despite these advances, a central challenge remains unresolved, namely achieving high-fidelity, physics-consistent designs at scale and in real time.

Classical topology optimization methods are predominantly iterative and gradient-driven and can be broadly classified into density-based and boundary-based formulations. Among these, density-based approaches such as the solid isotropic material with penalization (SIMP) method and related interpolation schemes remain the most widely adopted \cite{bendsoe1989optimal, yang1996stress, sigmund_topology_2013}. Complementing this class, level-set methods provide a boundary-based framework in which structural geometries are implicitly represented and evolved, typically through Hamilton--Jacobi-type equations governing boundary motion \cite{allaire2004structural, wang2003level}.

Both classes of methods rely on repeated cycles of finite element analysis (FEA), sensitivity evaluation, and design updates. In density-based formulations such as SIMP, the design variables correspond to element-wise material densities and are typically updated using gradient-based mathematical programming techniques, most notably the method of moving asymptotes (MMA) \cite{svanberg1987mma}. In contrast, level-set methods update the structural boundary by propagating an implicit surface using shape sensitivities, often coupled with constraint enforcement strategies such as augmented Lagrangian or penalty-based formulations.

While mathematically rigorous and capable of producing high-quality designs, both classes of methods remain computationally intensive and inherently iterative. The computational cost scales unfavorably with mesh resolution, multi-load scenarios, and high-dimensional design spaces. Moreover, convergence behavior depends on initialization and heuristic parameter choices, often requiring multiple restarts to obtain reliable solutions. Consequently, real-time topology optimization, critical for interactive design, rapid prototyping, and design-space exploration, remains largely out of reach \cite{mukherjee_accelerating_2021, behzadi_real-time_2021}.

A wide range of strategies has been proposed to mitigate the computational burden of classical topology optimization, including parallel computing, reduced-order modeling, and multi-resolution formulations \cite{borrvall_large-scale_2001, guest_reducing_2010, filomeno_coelho_model_2008, guo2014doing}. These approaches reduce the cost per iteration or improve scalability, but do not fundamentally alter the iterative nature of the optimization process. Other developments, such as isogeometric analysis, improve geometric representation and numerical accuracy within these iterative frameworks \cite{allaire2004structural}. Despite these advances, all such methods continue to rely on repeated PDE solves and iterative updates. This persistent dependence highlights a deeper limitation, namely the absence of a direct, non-iterative mapping from physical conditions to optimal topologies.

Machine learning (ML) has emerged as a promising paradigm to address this limitation by learning direct mappings from problem conditions to optimized designs, thereby reducing or eliminating repeated PDE solves. Early approaches include feedforward neural networks such as TouNN \cite{chandrasekhar_tounn_2021}, physics-informed neural networks \cite{jeong_physics-informed_2023}, and extensions to more complex settings such as multi-material design \cite{shishir_multimaterials_2024} and graph-based representations \cite{tabarraei2025graph}. Convolutional encoder--decoder architectures have also been widely explored due to their ability to exploit spatial locality in structured domains \cite{oh_deep_2019, senhora2022machine, zhang2025real}. More recently, neural implicit representations such as NITO \cite{nobari2024nito} have enabled resolution-independent topology generation within a learning-based framework. In addition, self-supervised latent-variable formulations, including hybrid quantum–classical approaches, have been proposed, where structural designs are parameterized in a low-dimensional latent space and optimized directly using physics-based objectives without relying on precomputed datasets \cite{tabarraei2025variational}.
Generative adversarial networks have further improved the quality of generated designs \cite{nie_topologygan_2021}, while diffusion-based models have achieved state-of-the-art structural fidelity by iteratively refining designs through learned denoising processes \cite{maze_diffusion_2022, giannone_diffusing_2023, zhang_research_2025, lutheran2025latent}. A comparison of representative ML-based approaches is summarized in Table \ref{table:intro_comparison}.

Despite these advances, a fundamental challenge remains. Direct prediction models enable fast inference but often struggle to enforce physical constraints and achieve high structural fidelity, while iterative generative approaches such as diffusion models achieve strong performance at the cost of substantial computational overhead during inference. Related developments in hybrid learning frameworks, including quantum-enhanced latent representations \cite{tabarraei2025variational}, further emphasize the importance of efficient representation and sampling in high-dimensional design spaces. Taken together, these observations point to a central challenge in ML-based topology optimization, namely developing models that can capture global, nonlocal interactions while maintaining computational efficiency at inference time.

From this perspective, topology optimization can be interpreted as learning a global operator that maps physics-informed inputs, such as boundary conditions, loads, and derived fields, to optimal material distributions. This mapping is inherently nonlocal, since the optimal placement of material depends on long-range interactions across the domain induced by equilibrium constraints. Capturing these interactions is essential for achieving both accuracy and generalization in non-iterative TO frameworks.

Transformers provide a natural architectural foundation for this problem. Originally developed for sequence modeling \cite{vaswani2023attentionneed}, and later extended to vision tasks via Vision Transformers (ViT) \cite{dosovitskiy2020image, zhai2022scalingvisiontransformers}, transformers employ self-attention mechanisms that explicitly model global interactions between all input tokens. Unlike convolutional architectures, which rely on local receptive fields and hierarchical aggregation, transformers directly capture long-range dependencies in a single layer. This property aligns closely with the global nature of structural mechanics, where loads, supports, and material distributions interact nonlocally to determine structural response. Recent works have demonstrated the effectiveness of transformer-based architectures for operator learning and PDE modeling \cite{cao2021choose, li2021fourier}, yet their application to topology optimization remains limited.

\begin{table*}
\caption{Comparison of representative machine learning approaches for topology optimization, highlighting trade-offs between physical fidelity and computational efficiency.}
\begin{tabular}{|m{2.8cm}|m{3.2cm}|m{3.6cm}|m{2.4cm}|m{2.4cm}|}
\hline
\rowcolor[HTML]{C0C0C0} 
\textbf{Method} & \textbf{Model Type} & \textbf{Key Idea} & \textbf{Fidelity} & \textbf{Inference Cost} \\
\hline
TouNN \cite{chandrasekhar_tounn_2021} 
& Neural Network 
& Per-instance training using compliance loss 
& High 
& High \\
\hline

TopologyGAN \cite{nie_topologygan_2021} 
& GAN 
& Direct generation conditioned on physics fields 
& Medium 
& Low \\
\hline

TopoDiff \cite{maze_diffusion_2022} 
& Diffusion Model 
& Iterative denoising guided by physics 
& High 
& Very High \\
\hline

NITO \cite{nobari2024nito} 
& Neural Implicit Representation 
& Continuous, resolution-independent topology learning 
& Medium--High 
& Medium \\
\hline

Hybrid Quantum ML \cite{tabarraei2025variational} 
& VQC + Decoder 
& Quantum-enhanced latent representation 
& High 
& High \\
\hline

\textbf{This Work} 
& Transformer (ViT) 
& Global attention-based operator learning 
& High 
& Low \\
\hline
\end{tabular}
\label{table:intro_comparison}
\end{table*}

In this work, we develop a transformer-based framework for topology optimization that directly learns the mapping from physics-informed input fields to optimized structural layouts. The proposed approach embeds boundary conditions, loading information, and volume fraction into a learnable conditioning token, while representing stress and strain energy fields as tokenized spatial inputs processed through a Vision Transformer. Unlike diffusion-based models, the proposed framework produces optimized topologies in a single forward pass, eliminating iterative inference while maintaining high structural fidelity. To ensure physical consistency and manufacturability, we introduce auxiliary loss functions that enforce volume constraints, load adherence, and structural connectivity. Furthermore, we extend the framework to dynamic loading conditions through frequency-domain encoding of time-dependent loads using fast Fourier transforms (FFT), combined with transfer learning to leverage large static datasets.

Taken together, this work advances topology optimization toward a non-iterative, operator-learning paradigm that integrates physics-informed modeling with modern attention-based architectures, enabling scalable and high-fidelity structural design.

\section{Topology Optimization}\label{sec:02}

Topology optimization determines the optimal material distribution within a prescribed design domain subject to loads, boundary conditions, and performance objectives. Depending on the application, objectives may include compliance minimization, thermal performance, fluid flow control, or more general multiphysics criteria. In structural settings, compliance is commonly used as a measure of structural performance, as it reflects the work done by external forces and is directly related to the stiffness of the system. In this work, we consider compliance minimization under a prescribed volume constraint.

\subsection{Static Structural Optimization}\label{static_topopt}

We adopt the standard density-based formulation using the SIMP method \cite{yang1996stress, sigmund_topology_2013}, in which the design domain is discretized into finite elements and each element is assigned a density variable $\rho_e \in [0,1]$. The topology optimization problem is formulated as

\begin{eqnarray}
\min_{\rho, \mathbf{u}} \quad C(\rho) = \mathbf{u}^T \mathbf{K}(\rho)\mathbf{u} \\
\text{s.t.} \quad \mathbf{K}(\rho)\mathbf{u} = \mathbf{f}, \\
\frac{V(\rho)}{V_0} \leq f, \\
0 \leq \rho_e \leq 1.
\end{eqnarray}

Here, $\mathbf{K}(\rho)$ is the global stiffness matrix, $\mathbf{u}$ is the displacement vector, and $\mathbf{f}$ is the applied load vector. The equilibrium constraint enforces the balance between internal and external forces, while the volume constraint limits material usage within the design domain.

Material properties are interpolated using the SIMP scheme
\begin{equation}
E_e(\rho_e) = E_{min} + \rho_e^p (E_0 - E_{min}),
\end{equation}
where $E_{min}$ is a small positive parameter introduced for numerical stability and $p$ is the penalization factor used to discourage intermediate densities.

The optimization is solved using gradient-based methods. The sensitivity of the compliance with respect to the design variables is given by
\begin{equation}
\frac{\partial C}{\partial \rho_e} = -p \rho_e^{p-1} \mathbf{u}_e^T \mathbf{K}_e \mathbf{u}_e,
\end{equation}
where $\mathbf{u}_e$ and $\mathbf{K}_e$ denote the element displacement vector and stiffness matrix, respectively. These sensitivities guide the update of the design variables within optimization algorithms such as the method of moving asymptotes  \cite{svanberg1987mma}.

The optimization proceeds iteratively until convergence, which is typically defined based on the relative change in compliance or a maximum number of iterations. The resulting density field $\rho$ defines the optimized topology.

\begin{figure*}
    \centering
    \includegraphics[width=1\linewidth]{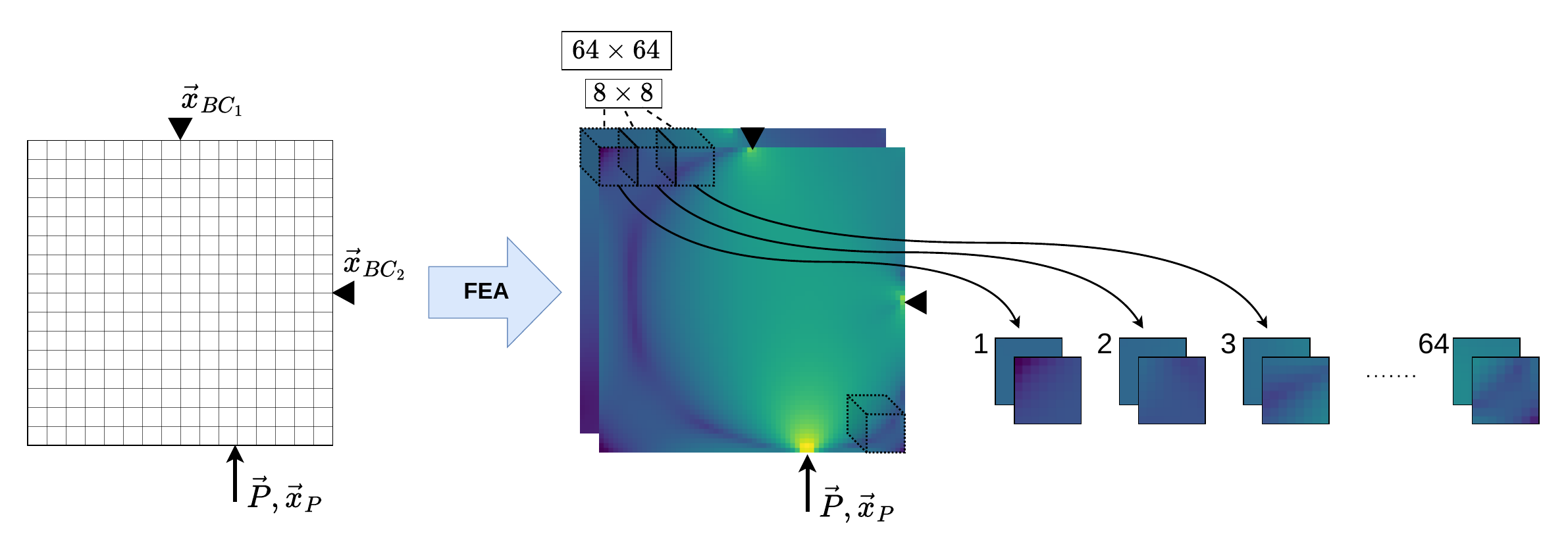}
    \caption{Processing for the inputs of the structural optimization problem for a given problem with boundary conditions $\vec{x}_{BC}$, load $\vec{P}$ at location $\vec{x}_P$, and volume fraction $f$. First, finite element analysis is used to convert the sparse inputs into dense images of the strain energy density and the von Mises stress. The fields of size $64 \times 64$ are then divided into a sequence of patches, in this example $8 \times 8$.}
    \label{fig:patches}
\end{figure*}
\subsection{Dynamic Structural Optimization}\label{dynamic_topopt}

The dynamic response of a structure under time-varying loads is governed by the finite element equation of motion, which accounts for inertia, damping, stiffness, and external forcing \cite{lee2018efficient, rong2000topology, zhao2016dynamic, lee2015nonlinear}. Unlike static loading, the response depends on inertia, damping, and the temporal characteristics of the excitation, and cannot, in general, be represented by an equivalent static load, thus requiring explicit treatment of time-dependent effects. The governing equation is given by
\begin{equation}
\mathbf{M}(\rho) \ddot{\mathbf{u}}(t) + \mathbf{C}(\rho) \dot{\mathbf{u}}(t) + \mathbf{K}(\rho) \mathbf{u}(t) = \mathbf{f}(t)
\end{equation}
where $\mathbf{M}(\rho)$, $\mathbf{C}(\rho)$, and $\mathbf{K}(\rho)$ denote the global mass, damping, and stiffness matrices, respectively, all dependent on the design field $\rho$. The vector $\mathbf{f}(t)$ represents the time-dependent load, and $\mathbf{u}(t)$ is the displacement response.

The element mass matrix is typically obtained from a consistent formulation
\begin{equation}
\mathbf{M}_e = \int_{\Omega_e} \rho_e \, \mathbf{N}^T \mathbf{N} \, d\Omega,
\end{equation}
although lumped mass approximations are often employed for computational efficiency, leading to a diagonal representation $\mathbf{M}_e = m_e \mathbf{I}$.

Damping is commonly modeled using Rayleigh damping,
\begin{equation}
\mathbf{C} = \alpha \mathbf{M} + \beta \mathbf{K},
\end{equation}
where $\alpha$ and $\beta$ are coefficients chosen to approximate desired modal damping behavior. Material properties remain design-dependent and are interpolated in a manner consistent with the SIMP formulation. For example, damping parameters may be expressed as
\begin{equation}
c(\rho_e) = c_{\min} + \rho_e^p (c_0 - c_{\min}).
\end{equation}

The objective in dynamic topology optimization is often defined in terms of dynamic compliance, given in the time domain by
\begin{equation}
C_{\text{dyn}}(\rho) = \int_0^T \mathbf{f}(t)^T \mathbf{u}(t)\, dt.
\end{equation}
The corresponding sensitivities with respect to the design variables are
\begin{equation}
\frac{\partial C_{\text{dyn}}}{\partial \rho_e} = 
\int_0^T 
\mathbf{u}_e(t)^T 
\frac{\partial \mathbf{K}_e}{\partial \rho_e}
\mathbf{u}_e(t)\, dt,
\end{equation}
which extend the static compliance sensitivities to the time-dependent setting.

In practice, the time domain is discretized into $N$ time steps, yielding
\begin{align}
C_{\text{dyn}}(\rho) &\approx \sum_{i=0}^N \mathbf{f}(t_i)^T \mathbf{u}(t_i)\, \Delta t, \\
\frac{\partial C_{\text{dyn}}}{\partial \rho_e} &\approx 
\sum_{i=0}^N 
\mathbf{u}_e(t_i)^T 
\frac{\partial \mathbf{K}_e}{\partial \rho_e}
\mathbf{u}_e(t_i)\, \Delta t.
\end{align}
These quantities are used within gradient-based optimization frameworks such as MMA \cite{svanberg1987mma}, with convergence assessed based on changes in the dynamic compliance.

Compared to static optimization, the dynamic formulation significantly increases computational cost, as the governing equations must be solved at each time step for every design iteration. This repeated time integration and sensitivity evaluation leads to substantial computational overhead, particularly for fine spatial and temporal discretizations. As a result, developing approaches that reduce or bypass this iterative time-dependent analysis is critical for enabling efficient dynamic topology optimization.

\section{Transformers}\label{sec:3}
Transformer architectures have been used across many applications, such as sequence modeling for natural language processing and spatial modeling for image processing. Unlike convolutional neural networks, which rely on learning local relationships in image data \cite{he2015deepresiduallearningimage}, transformers use a self-attention mechanism to model the interactions between all input elements of a sample \cite{vaswani2023attentionneed}. This allows them to model non-local relationships, which greatly improves their performance on problems that cannot be assumed to have exclusively local behavior. Transformers are therefore suited for problems that relate distant parts of the input domain, such as the distant load and boundary conditions seen in topology optimization problems.

In order to leverage transformer models for structural optimization, we first need to represent the input conditions as a sequence. In image processing, a common technique for implementing transformers is to patchify the input image, where the image is divided into non-overlapping image patches using a window of size $P \times P$. This patching process converts the spatial domain into a collection of smaller, fixed-size patches. For an input image with height $H$, width $W$ and number of channels $C_{ch}$, each patch has size $C_{ch}\times P\times P$. The total number of patches $N$ is then computed by $N = \frac{HW}{P^2}$. \cref{fig:patches} illustrates this patching process, demonstrating how the inputs for our structural optimization process are transformed into patches.

Each patch is then flattened into a vector and linearly projected into a token embedding space. Specifically, each flattened patch $x_i \in \mathbb{R}^{P^2 C_{ch}}$ is multiplied by a learnable projection matrix $E \in \mathbb{R}^{D \times P^2C_{ch}}$ where $D$ is the embedding dimension. The result of this projection is a vector $z_i^0 = Ex_i$, which is referred to as a 'token'. This process is applied to each patch with index $i \in \{1, 2, ... N\}$, mapping the entire image into a sequence of token embeddings $X \in \mathbb{R}^{N\times D}$. Because the same projection is applied to every patch, the resulting tokens lose information about their original locations in the input. Spatial relationships carry important information, so it is necessary to reintroduce positional information into the tokens. A simple method to accomplish this is to prepend a position index to the new representation based on its position in the full image.

Transformers utilize a multi-head self-attention mechanism, which weighs the importance of each input token relative to all other input tokens. For a sequence of token embeddings $X \in \mathbb{R}^{N\times D}$, the self-attention mechanism performs three primary projections to the data
\begin{eqnarray}
    Q = XW^Q; \qquad K = XW^K; \qquad V = XW^V 
\end{eqnarray}
This projection maps the input token sequence $X$ to three representations, namely the query $Q$, key $K$, and value $V$. This is a learnable projection via the weight matrices $W^Q$, $W^K$, and $W^V$, where each projection matrix lies in $\mathbb{R}^{D\times D}$. These projections are used for the computation of attention scores with scaled dot-product attention

\begin{equation}
    A_{ij} = \frac{\mathbf{q}_i^T \mathbf{k}_j}{\sqrt{d_k}}.
\end{equation}
The vectors $\mathbf{q}_i$ and $\mathbf{k}_j$ denote the $i$-th query and $j$-th key vectors, respectively, and $d_k$ is the dimensionality of the key vectors, used to scale the dot-product for numerical stability.
The resulting attention values are normalized by a softmax function across all tokens

\begin{equation}
    A_{ij} =  \frac{\exp(A_{ij})}{\sum_{j=1}^N \exp(A_{ij})}
\end{equation}
These normalized weights quantify how strongly token $i$ attends to token $j$, indicating the relative importance of information from token $j$ when updating token $i$. The updated representation of each token is computed as a weighted combination of the value vectors
\begin{equation}
    \mathbf{z}'_i = \sum_{j=1}^N A_{ij}\,\mathbf{v}_j.
\end{equation}
Here, $\mathbf{v}_j$ denotes the value vector associated with the $j$-th token, and $\mathbf{z}'_i$ represents the updated embedding of the $i$-th token after aggregating information from all tokens according to the attention weights. 

\begin{figure}
    \centering
    \includegraphics[clip, width=1\linewidth]{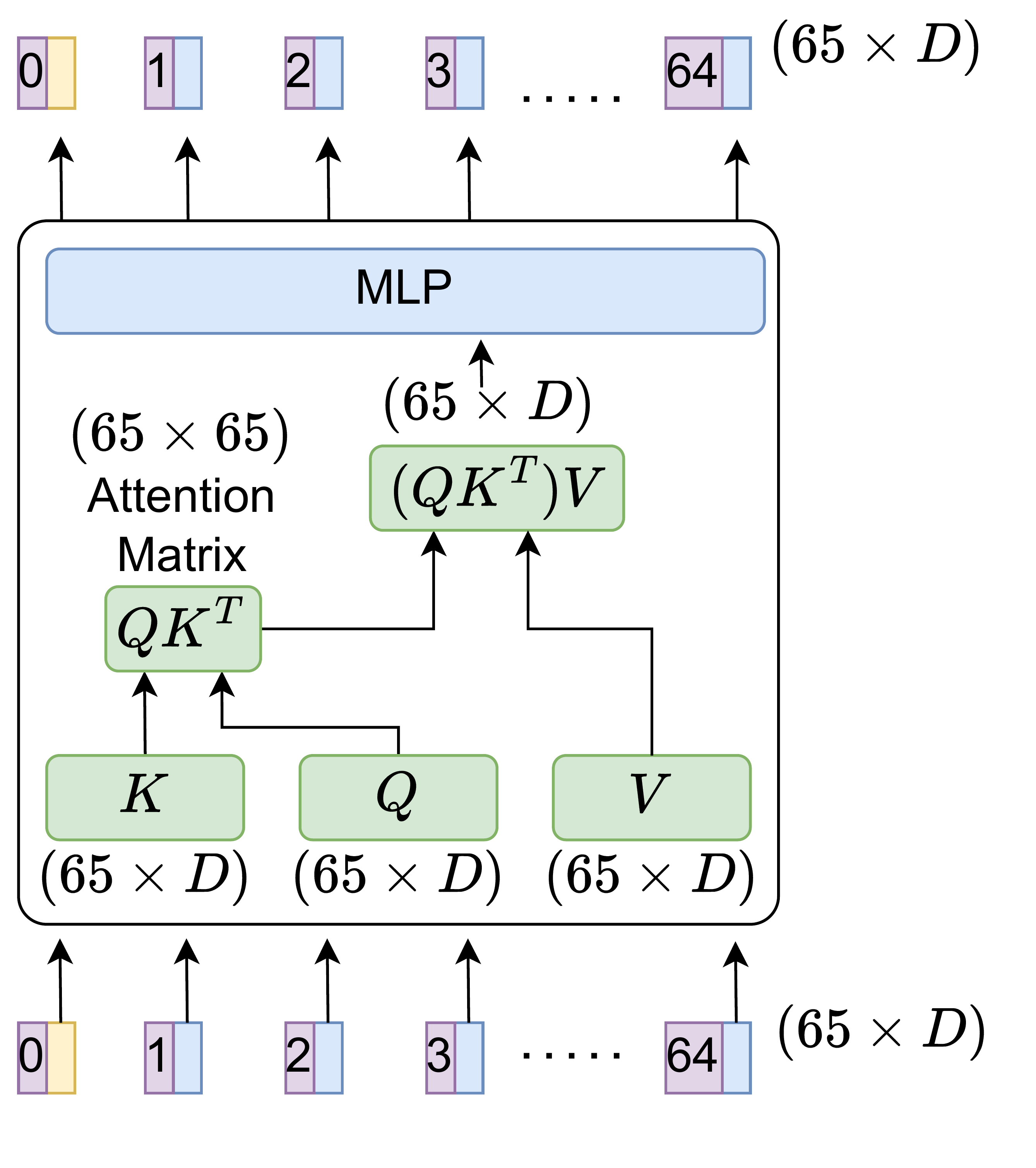}
    \caption{Transformer block architecture, which maps a set of input tokens $\mathbb{R}^{65\times D}$ into a similarly sized output. } 
    \label{fig:transformer_block}
\end{figure}

This operation enables each token to incorporate information from the entire domain, allowing the model to capture long-range spatial interactions that are essential for structural response. The updated token representations are then passed through a feed-forward multilayer perceptron (MLP), consisting of linear transformations and nonlinear activations. Residual connections are applied to stabilize training and preserve information across layers. The overall structure of a transformer block, including the attention and MLP components, is illustrated in \cref{fig:transformer_block}.

\begin{table}
    \begin{tabular}{|P{0.58in}|P{0.49in}|P{0.38in}|P{0.36in}|P{0.62in}|}\hline
        \rowcolor[HTML]{C0C0C0} 
        \textbf{Model} & \textbf{Hidden Dim ($D$)} & \textbf{Layers ($L$)} & \textbf{Heads ($h$)} & \textbf{Parameters (approx.)} \\
        \hline
        ViT-Tiny  & 192              & 12           & 3           & 5M                   \\
        ViT-Small & 384              & 12           & 6           & 22M                  \\
        ViT-Base  & 768              & 12           & 12          & 86M                  \\
        ViT-Large & 1024             & 24           & 16          & 307M                 \\
        ViT-Huge  & 1280             & 32           & 16          & 632M                
    \\ \hline\end{tabular}%

    \caption{Comparison of ViT sizes, including hidden dimension $D$, number of layers $L$, number of heads $h$, and total number of trainable parameters.}
    \label{table:ViTs}
\end{table}

\subsection{Vision Transformers}\label{sec:3a}
Vision Transformer (ViT) \cite{dosovitskiy2020image} is a deep learning architecture that leverages self-attention based transformer blocks for image processing, rather than natural language processing. ViTs rely on the patchify and position embedding process to map images into token sequences for transformer processing.

ViT models are typically characterized by three key hyperparameters, namely the embedding dimension $D$, the number of encoder layers $L$, and the number of attention heads $h$. These parameters control the representational capacity of the model as well as its computational cost. Smaller configurations, such as ViT-Tiny or ViT-Small, use reduced embedding dimensions and fewer attention heads, making them suitable for limited data settings. Larger configurations increase both depth and embedding size, enabling the model to capture more complex interactions at the expense of increased memory and training cost. A summary of representative ViT configurations is provided in \cref{table:ViTs} \cite{dosovitskiy2020image, zhai2022scalingvisiontransformers}.

The choice of model size depends strongly on the available data. Smaller models often generalize better in data-limited regimes, while larger models benefit from increased training data and computational resources, allowing them to learn richer and more expressive representations.

Beyond architectural scaling, an important training strategy for improving representation learning in vision transformers involves token masking. When trained as a masked autoencoder, a vision transformer exhibits improved efficiency, scalability, and generalization capacity \cite{he_masked_2021}. In this approach, a random subset $\mathcal{M} \subset \{1,\dots,N\}$ of patch tokens from an image sequence is selected for masking. Tokens corresponding to indices in $\mathcal{M}$ are replaced by a special learned embedding $\mathbf{z}_{\text{mask}}$, while unmasked tokens remain unchanged
\begin{equation}
\tilde{\mathbf{z}}_i = 
\begin{cases} 
    \mathbf{z}_i & i \notin \mathcal{M} \\ 
    \mathbf{z}_{\text{mask}} & i \in \mathcal{M} 
\end{cases}
\end{equation}
The network’s objective is then to reconstruct the original embeddings or pixel values corresponding to these masked patches using contextual information from unmasked tokens. This forces the model to learn meaningful global representations rather than relying solely on local cues. The reconstruction loss is typically formulated as mean squared error (MSE) over only the masked positions
\begin{equation}\label{mainLoss}
\mathcal{L}_{\text{mask}} = \frac{1}{|\mathcal{M}|} \sum_{i \in \mathcal{M}} \|\hat{\mathbf{x}}_i - \mathbf{x}_i\|^2
\end{equation}
where $\hat{\mathbf{x}}_i$ denotes the reconstructed patch embedding and $\mathbf{x}_i$ is the ground truth embedding for patch $i$. This masking-based pretraining strategy has been shown to significantly improve downstream performance by encouraging robust feature extraction from incomplete visual information \cite{dosovitskiy2020image}. This strategy also encourages a reliance on multiple tokens across the image, preventing the model from overfitting on small subsets of the image domain.

\begin{figure*}
    \centering
    \includegraphics[trim={0mm 3mm 2mm 2mm}, clip, width=1\linewidth]{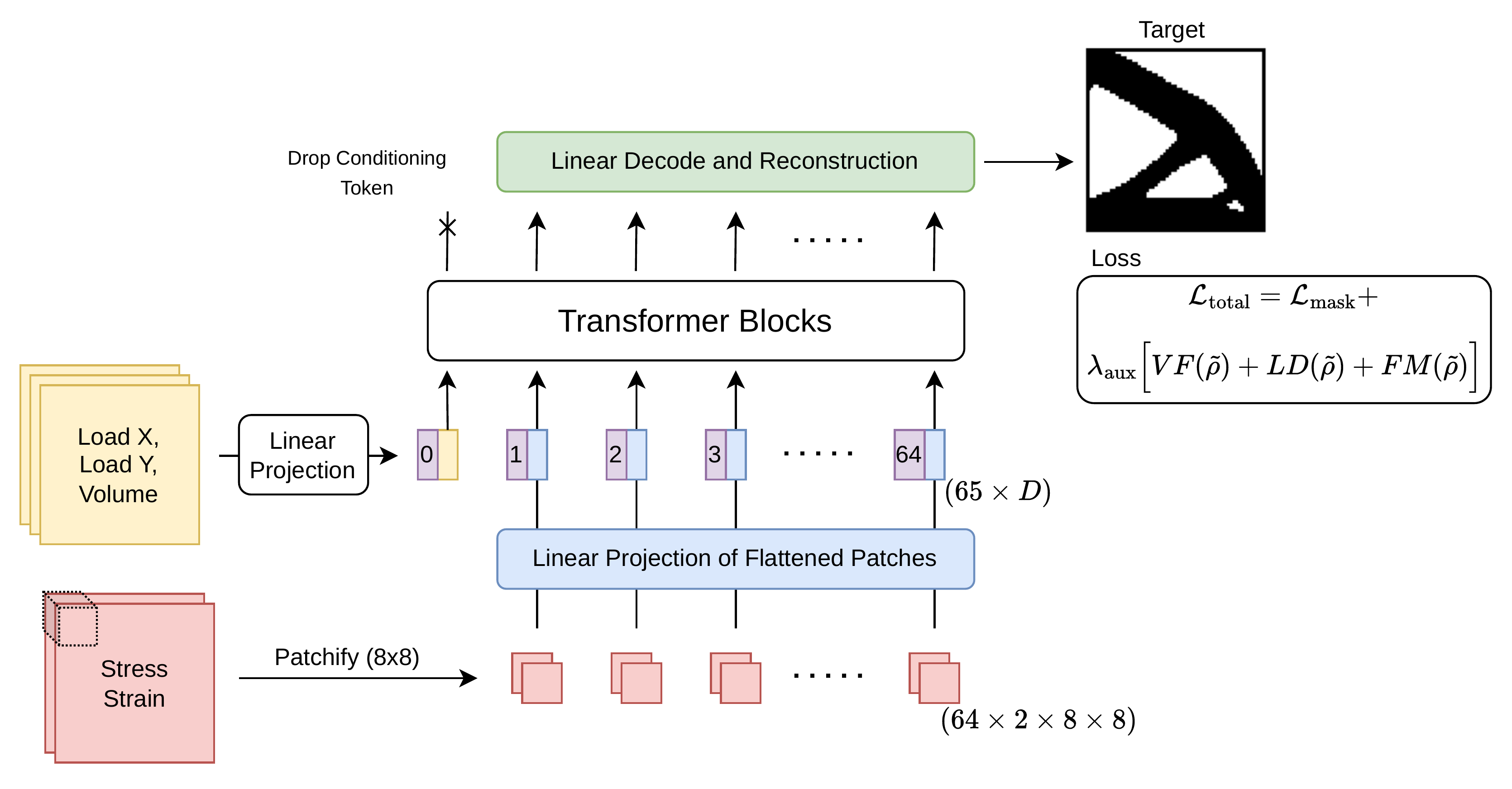}
    \caption{Full proposed transformer architecture, including a separate treatment of the loading conditions and volume fraction from the stress and strain fields. Example patchification uses an $8\times8$ patch size, producing 64 tokens, although $4\times4$ and $2\times2$ patch sizes are also studied, yielding 256 and 1024 tokens respectively.}
    \label{fig:transformer_full}
\end{figure*}

\section{Methodology}\label{sec:4}

\subsection{Dataset}\label{sec:4a}
Each sample in the dataset represents a unique structural optimization problem, with load, volume fraction, and boundary conditions. From these inputs, the optimized topology of the sample is calculated using the SIMP optimization scheme, as described in \cref{sec:02}. The topology is stored as a binary grayscale image of the design domain, where pixels with values of 1 represent material regions and pixels with values of 0 represent void.

In addition to the optimized topology and problem definitions, each sample has corresponding physical field representations. The physical fields capture the dense spatial relationship between the boundary conditions and loading information used during optimization. Specifically, these are represented as distributions of the strain energy density and von Mises stress, computed on the initial unoptimized design domain. These fields are calculated this way so that all input quantities are known prior to sampling. To maintain consistency across samples, these fields are normalized to the range $[0,1]$.

The von Mises stress is computed from the stress tensor as
\begin{equation}
\sigma_{\text{vm}} = \sqrt{\tfrac{3}{2}\,\mathbf{s}:\mathbf{s}},
\end{equation}
where $\mathbf{s}$ is the deviatoric stress tensor. The strain energy density is given by
\begin{equation}
w = \tfrac{1}{2}\boldsymbol{\sigma} : \boldsymbol{\varepsilon}.
\end{equation}
Both quantities are evaluated at element centers and used as scalar input fields.

The dataset is generated on a $64 \times 64$ grid of square finite elements. External loads are applied at randomly selected boundary elements with unit magnitude, and their direction is sampled from six evenly spaced angles between $0^\circ$ and $360^\circ$. Boundary conditions are defined by selecting between one and four constraints from a set of predefined node groups, including corner nodes, midpoints, and edge segments, with all constrained nodes fixed in both spatial directions. The volume fraction is uniformly sampled between $30\%$ and $50\%$.

The von Mises stress and strain energy density are evaluated at the element centers of the initial unoptimized domain during the pre-processing of the dataset. The finite element analysis assumes a unit linear elastic modulus and a Poisson's ratio of 0.3. These values only determine the scaling of the fields, rather than their relative distributions. In total, 30,000 samples were generated, with further dataset diversity being achieved from rotating and mirroring the samples. These transformations expand the effective training set by a factor of eight relative to the original dataset. This approach improves the sample efficiency and encourages the model to learn rotationally and reflection-invariant features, reducing overfitting and improving generalization without additional labeled data.

\subsection{Architecture}\label{sec:4b}
The proposed model architecture builds upon the Vision Transformer (ViT) framework described in \sref{sec:3a}, adapted for physics-informed inputs. An overview of the architecture and the data flow through the model is shown in \cref{fig:transformer_full}.
The model input consists of a two-channel image defined over a $64 \times 64$ domain, where the channels correspond to strain energy density and von Mises stress. Unlike conventional image-based models that operate on three-channel RGB inputs, these channels represent physically meaningful quantities that characterize the structural response.
To convert the spatial input into a form suitable for transformer processing, the domain is partitioned into non-overlapping patches. Each patch corresponds to a local region of the input fields and is treated as an individual token. As illustrated in \cref{fig:patches}, this patchification step preserves local spatial structure while enabling the representation of the domain as a sequence of tokens for subsequent processing by the transformer.

 The inputs consist of spatial fields of strain energy density and von Mises stress, which are treated as a multi-channel image and partitioned into patch tokens as described earlier. In addition to these spatial inputs, global problem parameters including volume fraction, load coordinates, load vector components, and boundary conditions are incorporated through a conditioning token.
This conditioning token is constructed by concatenating the global parameters and projecting them through a neural network into the same embedding dimension as the patch tokens. This enables the model to integrate global information with spatially distributed features through the attention mechanism. The resulting token sequence consists of one conditioning token and a set of embedded patch tokens.

All tokens, including both the conditioning token and the patch tokens derived from the stress and strain fields, are used to compute the query $Q$, key $K$, and value $V$ matrices in the self-attention mechanism. The attention output is then processed by a feed-forward multilayer perceptron (MLP), forming a standard transformer block. This process is repeated across $L$ stacked transformer layers.

After the transformer layers, the conditioning token is discarded, and the remaining tokens are mapped back to the spatial domain to reconstruct a single-channel output. This output represents the predicted density field over the design domain. A single forward pass of the model produces a topology conditioned on both the spatial physics fields and the global loading and boundary conditions.
The model is trained by minimizing the pixel-wise difference between the predicted density field and the ground truth topology.

\subsection{Auxiliary Losses}\label{sec:4c}
To promote physically consistent and manufacturable designs, the model is trained with auxiliary loss terms in addition to the primary loss in \eref{mainLoss}. These losses enforce structural constraints that are not captured by pixel-wise similarity, guiding the optimization toward globally coherent, load-bearing designs without explicitly minimizing compliance.

The first auxiliary term enforces consistency with the prescribed material usage through a volume fraction loss. Given a target volume fraction $f$, the predicted topology $\tilde{\rho}$ is penalized based on the deviation of its total material volume from the desired value,
\begin{equation}
VF(\tilde{\rho}) = \left| f - \frac{V(\tilde{\rho})}{V_0} \right|,
\end{equation}
where $V(\tilde{\rho})$ denotes the total volume of material in the predicted design and $V_0$ is the volume of the design domain. This term ensures that the generated topologies satisfy global material constraints.

The second auxiliary term addresses the placement of material relative to the applied loads. The load discrepancy loss penalizes designs that fail to allocate sufficient material in regions where loads are applied,
\begin{equation}
LD(\tilde{\rho}) = 1 - \sum_{e=1}^{N_e}\sqrt{(\tilde \rho_e F^x_e)^2 + (\tilde \rho_e F^y_e)^2},
\end{equation}
where $F_e$ is the load vector at element $e$. This loss encourages the model to place material in load-bearing regions, promoting physically meaningful structural layouts.

While the volume fraction and load discrepancy losses enforce global constraints and local load support, they do not guarantee that the resulting topology forms a single connected structure. In data-driven settings, models trained with pixel-wise losses may produce visually plausible designs that contain disconnected components or floating material, which do not contribute to load transfer and are inherently suboptimal.

To address this limitation, we introduce a differentiable floating-material loss, $FM(\tilde{\rho})$, which explicitly penalizes disconnected regions in the predicted topology. Unlike traditional connectivity checks based on graph traversal, which are non-differentiable, the proposed formulation provides a continuous approximation of global connectivity that can be integrated directly into gradient-based training.

\begin{algorithm}[t]
\caption{Differentiable floating-material (connectivity) loss}
\label{alg:fm-cont}
\begin{algorithmic}[1]
\Require Continuous density $\tilde\rho \in [0,1]^{H \times W}$; max iterations $T$; tolerance $\tau$; scale $\alpha$, shift $\beta$
\State $(i^\star,j^\star) \gets$ first argmax of $\tilde\rho$ 
\State Initialize $H \gets \mathbf{0}$, then $H_{i^\star j^\star} \gets 1$
\State Let $\mathbf{S} = \tfrac{1}{2}\left[\begin{smallmatrix}0&1&0\\1&1&1\\0&1&0\end{smallmatrix}\right]$
\Repeat[for $t = 1,\ldots,T$]
    \State $H_{\mathrm{prev}} \gets H$
    \State $\tilde{H}_{\mathrm{neigh}} \gets H * \mathbf{S}$ \Comment{2D convolution,}
    \State $A \gets \sigma\bigl(\alpha\,(\tilde{H}_{\mathrm{neigh}} - \beta)\bigr)$ \Comment{$\sigma$ = logistic}
    \State $H \gets \max\bigl(H,\, A \odot \tilde\rho\bigr)$ \Comment{element-wise}
\Until{$\|H - H_{\mathrm{prev}}\|_\infty < \tau$ or $t = T$} \Comment{$\tau = 10^{-4}$}
\State $FM(\tilde{\rho}) \gets \displaystyle\sum_{(i,j) \neq (i^\star,j^\star)} (\tilde{\rho}_{ij} - H_{ij})\, \tilde{\rho}_{ij}$
\Ensure $FM(\tilde{\rho})$
\end{algorithmic}
\end{algorithm}

The method, detailed in \cref{alg:fm-cont}, begins by selecting a seed at the location of maximum density and constructing a connectivity map through iterative propagation. At each iteration, connectivity is expanded to neighboring elements using a convolutional operator defined by a fixed kernel $\mathbf{S}$, followed by a smooth thresholding step implemented via a sigmoid function with parameters $\alpha$ and $\beta$. The propagation is restricted to regions containing material through element-wise multiplication with the density field, ensuring that connectivity can only extend through physically meaningful paths.

This process can be interpreted as a differentiable flood-fill operation, where connectivity propagates from the primary structural component through adjacent material. The floating material loss then measures the discrepancy between the predicted density field and the resulting connectivity map, effectively penalizing any material that is not reachable from the main structure,
\begin{equation}
FM(\tilde{\rho}) = \sum_{(i,j)\neq(i^*,j^*)} (\tilde{\rho}_{ij} - H_{ij})\,\tilde{\rho}_{ij}.
\end{equation}

If the predicted topology is fully connected, the connectivity map coincides with the density field and the loss vanishes. In contrast, disconnected components remain unvisited during propagation and are directly penalized. This introduces a strong inductive bias toward single, connected, load-bearing structures and suppresses the formation of isolated material regions.

In practice, this loss produces stable gradients that discourage floating material and sharpen topology boundaries, leading to designs that are both physically meaningful and more consistent with classical topology optimization solutions. To ensure balanced contributions, all auxiliary losses are normalized by the number of elements.

The overall training objective is defined as a weighted combination of the primary loss and the auxiliary losses
\begin{equation}
\mathcal{L}_{\text{total}} = \mathcal{L}_{\text{mask}} 
+ \lambda_{\text{aux}} \, \big[VF(\tilde{\rho}) 
+ \, LD(\tilde{\rho}) 
+ \, FM(\tilde{\rho})\big],
\end{equation}
where $\lambda_{\text{aux}}$ is the weighting coefficient that controls the relative importance of the auxiliary losses. For this work, we use an auxiliary loss weighting of $\lambda_{\text{aux}} = 0.075$.


\begin{figure}[bh]
    \centering
    
    \begin{subfigure}[b]{\linewidth}
        \caption{\label{fig:loads:subA}}
        \includegraphics[width=\linewidth, trim={0cm 12.4cm 0cm 0cm}, clip]{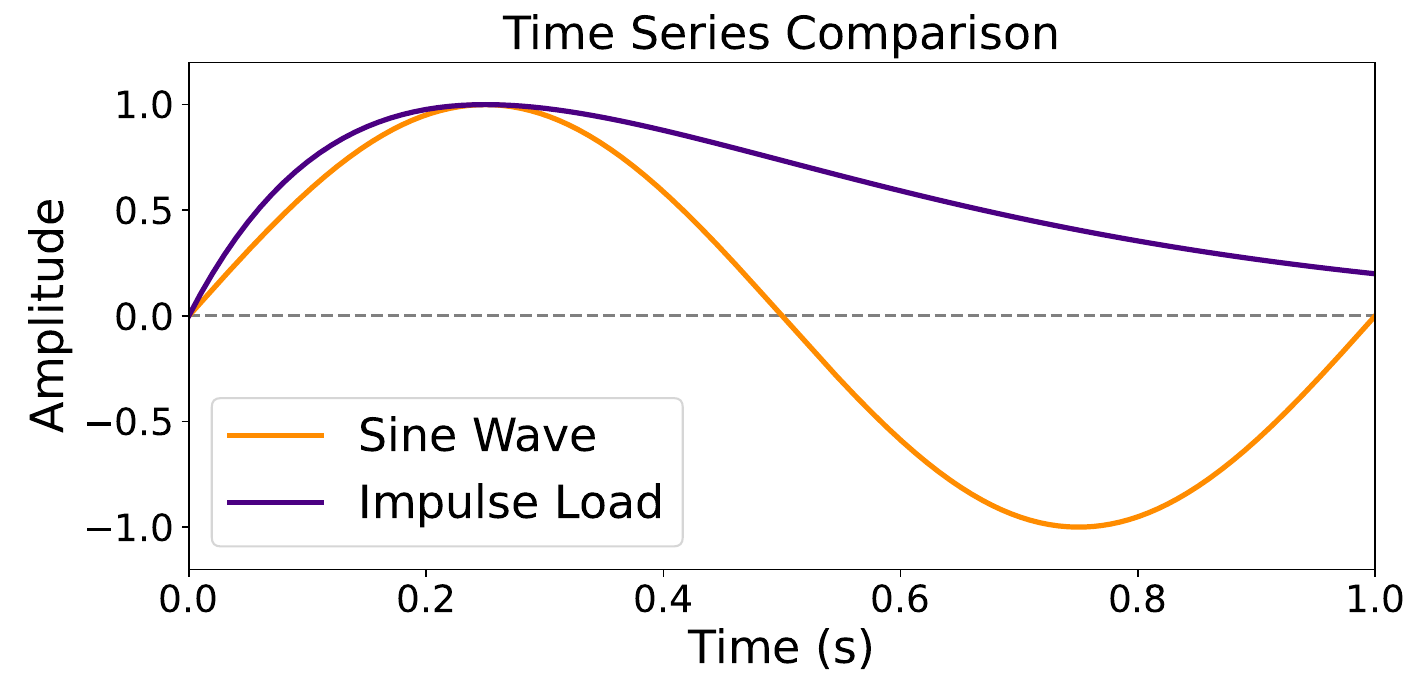}
    \end{subfigure}
    
    \begin{subfigure}[b]{\linewidth}
        \caption{\label{fig:loads:subB}}
        \includegraphics[width=\linewidth, trim={0cm 0cm 0cm 12.4cm}, clip]{fft_comparison_highres.pdf}
    \end{subfigure}
    
    \caption{Plot of the time sequence (a) and FFT amplitude transformation (b) for the two loads used in the dynamic dataset, sine (orange) and impulse (indigo).}
    \label{fig:loads}
\end{figure}


\subsection{Transfer Learning for Dynamic Topology Optimization}\label{sec:4d}

Dynamic topology optimization presents a significantly more challenging problem than its static counterpart, as the structural response depends on the temporal evolution of applied loads. Consequently, generating dynamic optimization data is substantially more computationally expensive, resulting in a smaller dataset of approximately 6{,}000 samples compared to 30{,}000 samples for the static case.

In this work, two representative time-dependent load functions are considered: a sinusoidal load and an impulse load, defined over the time interval $t \in [0,1]$,
\begin{align}
    s(t) = \sin(2\pi t), \label{eqn:sine} \\
    i(t) = \frac{t}{0.25} e^{-\tfrac{t}{0.25} + 1}. \label{eqn:impulse}
\end{align}
The corresponding time histories and their frequency-domain representations are shown in \cref{fig:loads}.

To address the limited size of the dynamic dataset, a transfer learning framework is employed. The transformer model is first trained on the larger static dataset, where it learns general mappings between physics-informed inputs and optimal topologies. This pre-trained model is then fine-tuned on the dynamic dataset. Since many geometric and structural features are shared between static and dynamic problems, this approach significantly reduces training cost while improving generalization and mitigating overfitting.

The spatial input channels, consisting of strain energy density and von Mises stress, are retained in the same manner as in the static formulation. These fields are computed on the initial, unoptimized design domain using linear elastic finite element analysis under a unit load configuration, and therefore provide a static characterization of the load distribution. In the dynamic setting, this serves as a baseline spatial approximation, while the time-dependent effects of the loading are incorporated separately through the conditioning token.

To account for time-dependent loading, the conditioning token is augmented with frequency-domain features derived from the applied load. The load signal is transformed using a discrete Fourier transform, yielding a set of frequency coefficients. The amplitudes of the first ten frequency components, $\omega_1, \dots, \omega_{10}$, are extracted and concatenated with the volume fraction, load location, and load vector components. This feature vector is then projected into the embedding space of the conditioning token.
This formulation enables the incorporation of temporal information without modifying the transformer architecture. The model continues to process spatial fields through patch tokens, while the conditioning token encodes global and temporal information. As a result, the pre-trained model can be efficiently adapted to dynamic problems with minimal retraining.

Only the first ten frequency components are retained, as the amplitude spectrum of the impulse load exhibits a rapid decay, indicating that the signal is strongly dominated by low-frequency content. In particular, the magnitude of the frequency components decreases by more than an order of magnitude within the first ten modes, while higher-frequency components contribute only a small fraction of the total spectral energy, as shown in \cref{fig:loads}.
From a structural dynamics perspective, this behavior is consistent with the response of elastic systems, where low-frequency excitation governs the global deformation and load transfer, while higher-frequency components are attenuated by inertial and damping effects and primarily influence localized, high-frequency vibrations. As a result, the topology optimization outcome is largely determined by the dominant low-frequency content of the loading.
Retaining only the first ten frequency components therefore captures the essential temporal characteristics of the load while avoiding unnecessary input dimensionality. Including higher-frequency components would introduce additional features with minimal physical relevance and limited representation in the dataset, potentially increasing model complexity without improving predictive performance.

This transfer learning framework allows the model to generalize from static to dynamic topology optimization, capturing both spatial and temporal characteristics of the loading while maintaining computational efficiency and robustness.

\section{Results and Analysis}\label{sec:5}
Each model is trained for 800{,}000 iterations with a batch size of 256. The dataset is split into 90\% for training and 10\% for validation. Model performance is evaluated using six metrics: compliance error, proportion of samples with compliance error exceeding 30\%, median compliance error, volume fraction error, load discrepancy, and floating material.

Compliance error is defined as the absolute relative difference between the structural compliance of the predicted topology and that of the ground truth
\begin{equation}
\text{Error}_{\text{comp}} = \frac{|C_{\text{pred}} - C_{\text{true}}|}{C_{\text{true}}}.
\end{equation}
The average compliance error is computed over all samples, including those with large deviations (i.e., errors exceeding 30\%), to provide a comprehensive measure of model performance.  To ensure numerical stability, an upper threshold of 100\% is imposed to exclude extreme outliers associated with disconnected or near-void structures, where compliance becomes excessively large and no longer provides a meaningful physical measure.
Notably, only a very small fraction of samples fall into this high-error regime.

The proportion of samples with compliance error greater than 30\% is also reported as a measure of model reliability, capturing the frequency of significant deviations from the ground truth. Median compliance error is computed after excluding these high-error (failed) samples, providing a robust estimate of model accuracy for well-performing predictions.

In addition to compliance-based metrics, the auxiliary quantities, volume fraction error, load discrepancy, and floating material, are evaluated directly to assess the physical consistency of the predicted topologies.
All metrics are computed on a fixed validation subset of 500 samples. The same subset is used for all models to ensure a consistent and unbiased comparison.

\subsection{Static Model Results}\label{static_res}

For the static optimization study, five Vision Transformer (ViT) architectures—Tiny, Small, Base, Large, and Huge—are first evaluated using a patch size of $P=8$, following standard ViT configurations. This setting serves as a baseline to assess the effect of model capacity on topology generation. The results are summarized in \cref{table:static} and compared with GAN-based (TopologyGAN) and diffusion-based (TopoDiff) approaches.

\begin{table*}
\centering
\begin{tabular}{|l|l|l|l|l|l|l|l|}
\hline
\rowcolor[HTML]{C0C0C0} 
\textbf{Metric}                    & \textbf{TopologyGAN} & \textbf{TopoDiff}&\textbf{Tiny} & \textbf{Small} & \textbf{Base} & \textbf{Large}& \textbf{Huge} \\
\hline
Compliance Error (\%)              & 48.51             &4.39&   4.75&   6.07           & 3.67    & 10.68  & 8.00\\
Compliance Error Above   30\% (\%) & 10.11             &0.83&   6.40&   5.60           & 10.80   & 31.80  & 21.00\\
Median Compliance Error   (\%)     & 2.06              &2.56&   1.55&   2.92           & 0.91    & 2.63   & 2.03\\
Volume Fraction Error (\%)         & 11.87             &1.85&   3.00&   4.77           & 4.03    & 26.91  & 7.22\\
Load Discrepancy (\%)              & 0.0               &0.0&    0.60&   1.20           & 5.4     & 3.6    & 3.80\\
Floating Material (\%)             & 46.78             &5.54&   14.20& 17.20           & 16.60   & 44.00  &  43.20\\
\hline       
\end{tabular}%
\caption{Analysis metrics for the static topology optimization dataset with the ViT Tiny, Small, Base, Large, and Huge models with the base patch size of 8. Results for TopologyGAN \cite{nie_topologygan_2021} and TopoDiff \cite{maze_diffusion_2022} are reference values from the TopoDiff paper.}
\label{table:static}
\end{table*}

\begin{table*}
\centering

\begin{tabular}{|l|l|l|l|l|l|l|l|l}
\hline
\rowcolor[HTML]{C0C0C0} 
\textbf{Metric}                    &\textbf{Tiny-2} & \textbf{Tiny-4} & \textbf{Tiny-8} & \textbf{Small-2} & \textbf{Small-4} & \textbf{Small-8}\\
\hline
Compliance Error (\%)              &   3.14  &   1.94  &4.75  &  3.70     &  \textbf{1.86} & 6.07 \\
Compliance Error Above   30\% (\%) &   3.40  &   3.40  &6.40  &  5.20     &  \textbf{2.20} & 5.60 \\
Median Compliance Error   (\%)     &   0.49  &   0.43  &1.55  &  0.93     &  \textbf{0.32} & 2.92 \\
Volume Fraction Error (\%)         &   1.51  &   1.39  &3.00  &  2.74     &  \textbf{1.27} & 4.77 \\
Load Discrepancy (\%)              &   0.00  &   1.00  &0.60  &  0.00     &  \textbf{0.00} & 1.20 \\
Floating Material (\%)             &  10.60  &   8.20  &14.20 & 26.00     &  \textbf{6.60} & 17.20\\
\hline       
\end{tabular}%
\caption{Analysis metrics for the static topology optimization dataset with the ViT Tiny and Small models, with patch size $P=2$, $P=4$, and $P=8$.}
\label{table:static_patch}
\end{table*}

Among the ViT models, the Base architecture achieves the lowest compliance error (3.67\%) and median compliance error (0.91\%), indicating strong predictive accuracy for well-formed structures. However, this improvement comes at the cost of robustness, with 10.8\% of samples exceeding 30\% compliance error. In contrast, the Small model achieves the lowest failure rate (5.6\%) while maintaining competitive compliance performance, suggesting a more favorable balance between accuracy and reliability.

As model size increases beyond the Base configuration, performance deteriorates significantly. The Large and Huge models exhibit higher compliance errors, increased failure rates, and poor volume fraction tracking. For example, the Large model reaches a failure rate of 31.8\% and a volume fraction error of 26.91\%. This behavior is consistent with overfitting, as the dataset of 30{,}000 samples is insufficient to effectively train the substantially larger parameter counts of these architectures. These results highlight the importance of matching model capacity to dataset size in topology optimization tasks.

Despite achieving reasonable compliance performance, all $P=8$ configurations exhibit noticeable float

\noindent\begin{minipage}{\linewidth}
    \includegraphics[width=1\linewidth, trim={0cm, 0.3cm, 0cm, 0.2cm}, clip]{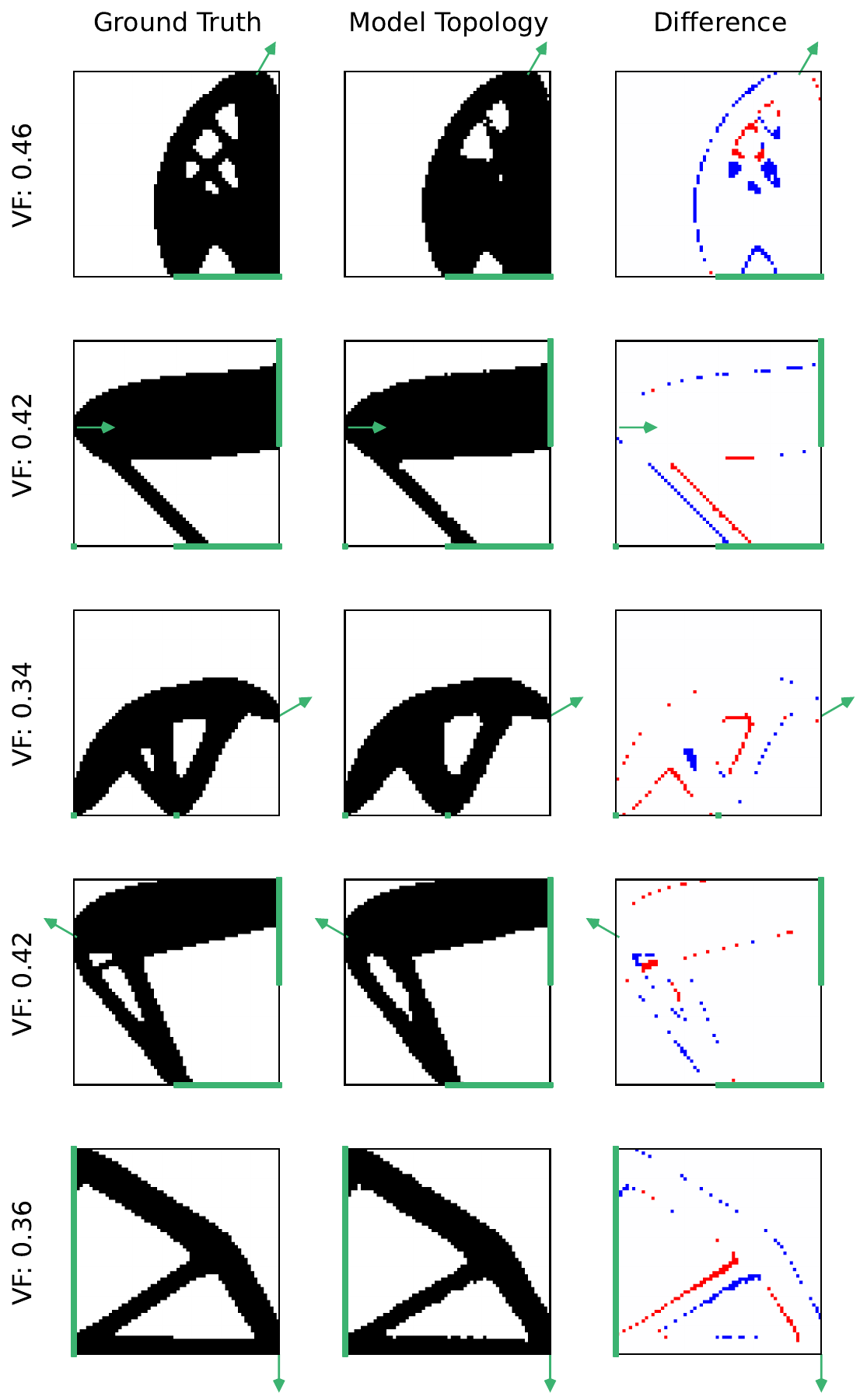}
    \captionof{figure}{Validation samples from the ViT Small model. From left to right, the columns depict the ground truth topology, model's predicted topology, and difference. Red indicates the model missing material while blue indicates excess material. Green sections indicate fixed boundary conditions and arrows indicate applied load. The bottom sample depicts the standard cantilevered beam problem with a fixed left wall and a downward point load in the bottom right corner.} 
    \label{fig:vitsmallsamples}
    \vspace{0.5cm}
\end{minipage}

\noindent-ing material, indicating limitations in capturing global structural connectivity. While local features, such as placing material near loads and supports, are consistently learned, enforcing long-range connectivity remains challenging at this resolution. Notably, even the best-performing ViT models exhibit higher floating material than the diffusion-based reference TopoDiff, suggesting that the observed connectivity limitations are not primarily due to model capacity, but rather to insufficient spatial resolution.

To investigate this further, we perform an ablation study on patch size using $P=4$ and $P=2$, focusing on the Tiny and Small architectures. Larger models are not included in this study, as the $P=8$ results already demonstrate that increasing model size beyond this range does not improve performance and leads to instability. Restricting the ablation to smaller models allows us to isolate the effect of spatial resolution without confounding it with overparameterization.

These ablation models retain the same architecture as the $P=8$ configurations, with the primary difference being an increase in the number of tokens as patch size decreases. The total number of parameters remains approximately constant, as projection weights are shared across tokens. The results are presented in \cref{table:static_patch}.

Reducing the patch size from $P=8$ to $P=4$ leads to substantial improvements across all metrics. For the ViT-Tiny model, the compliance error decreases from 4.75\% to 1.94\%, while floating material is reduced from 14.20\% to 8.20\%. Similar improvements are observed for the ViT-Small model, where the Small-$P=4$ configuration achieves the best overall performance, with a compliance error of 1.86\%, median error of 0.32\%, and floating material of 6.60\%.

However, further reducing the patch size to $P=2$ does not yield additional improvements and, in some cases, degrades performance. While $P=2$ increases spatial resolution, it also substantially increases the number of tokens, making the learning problem more difficult and reducing the model’s ability to capture coherent global structures. This suggests that excessively fine discretization introduces noise and training instability without improving structural fidelity.

These results indicate the presence of an optimal patch size, where sufficient spatial resolution is achieved without over-fragmenting the representation. In this study, $P=4$ provides the best balance, enabling the model to capture long-range connectivity while maintaining stable and efficient learning.

Overall, these findings demonstrate that the connectivity limitations observed at $P=8$ are primarily resolution-driven rather than architecture-driven. When sufficient spatial resolution is provided, the transformer framework produces high-quality, physically consistent topologies with low compliance error and significantly reduced floating material. 

Representative predictions from the ViT-Small model are shown in \cref{fig:vitsmallsamples}, illustrating the ground truth topology, the predicted topology, and their difference.

\subsection{Physical Consistency}
To further assess the physical fidelity of the generated topologies, we compare the peak stress and peak strain values of the predicted structures against those of the ground truth across 500 validation samples. The resulting scatter plots are shown in \cref{fig:stressstrain_stats}, where each point represents a single sample. A reference line $y=x$ is included to indicate perfect agreement between the model and ground truth.

The predicted stress values exhibit strong agreement with the ground truth, with most samples tightly clustered around the $y=x$ line. This indicates that the model successfully captures the primary load paths and stress distributions governing structural behavior. A number of samples fall below the reference line, suggesting that in some cases the model produces structures with improved (lower) peak stress relative to the ground truth.

The strain results follow a similar trend, although with slightly greater dispersion. While the majority of samples remain close to the $y=x$ line, there is a mild tendency for the predicted topologies to overestimate peak strain. This is consistent with the regression slopes reported in \cref{table:ss_stats}, where the slope for strain exceeds unity. Physically, this behavior suggests that the generated topologies may exhibit slightly higher local deformation, even when overall stress distributions are well captured.

\begin{figure}
        \centering
    \includegraphics[width=1\linewidth]{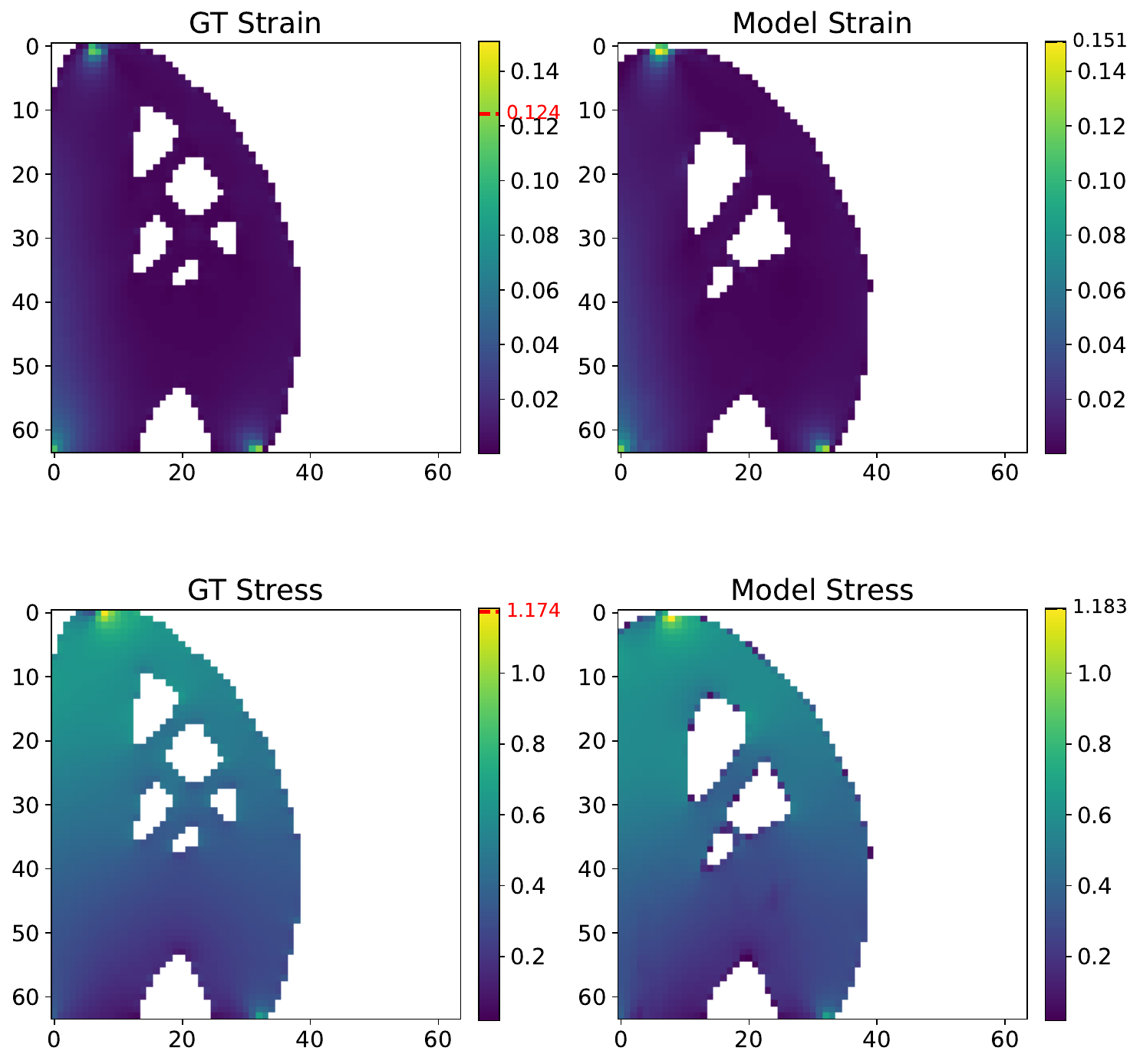}
    \caption{Stress and strain over an optimized sample after loading. The ground truth stress and strain are shown on the left. The model topology stress and strain are shown on the right. The maximum stress and strain are denoted on the color bar for the plots, with maximum values denoted. Samples are from the ViT-Small-8 model.}
    \label{fig:stressstrain}
\end{figure}

These observations are further illustrated in \cref{fig:stressstrain}, where the predicted topology reproduces the overall stress distribution of the ground truth, particularly along the primary load paths. The maximum von Mises stress is comparable between the two structures, while the predicted topology exhibits a slightly higher peak strain. This localized increase in strain is consistent with the trends observed in \cref{fig:stressstrain_stats} and suggests minor differences in stiffness distribution rather than a fundamental change in structural behavior.

\subsection{Post-Processing}

Additional post-processing can be applied to improve the quality of the generated topologies. Using the floating material loss metric $FM(\tilde{\rho})$, samples containing disconnected or non-load-bearing regions can be identi-

\noindent\begin{minipage}{\linewidth}
\centering
    \includegraphics[width=0.9\linewidth, trim={10mm 8mm 20mm 0mm}]{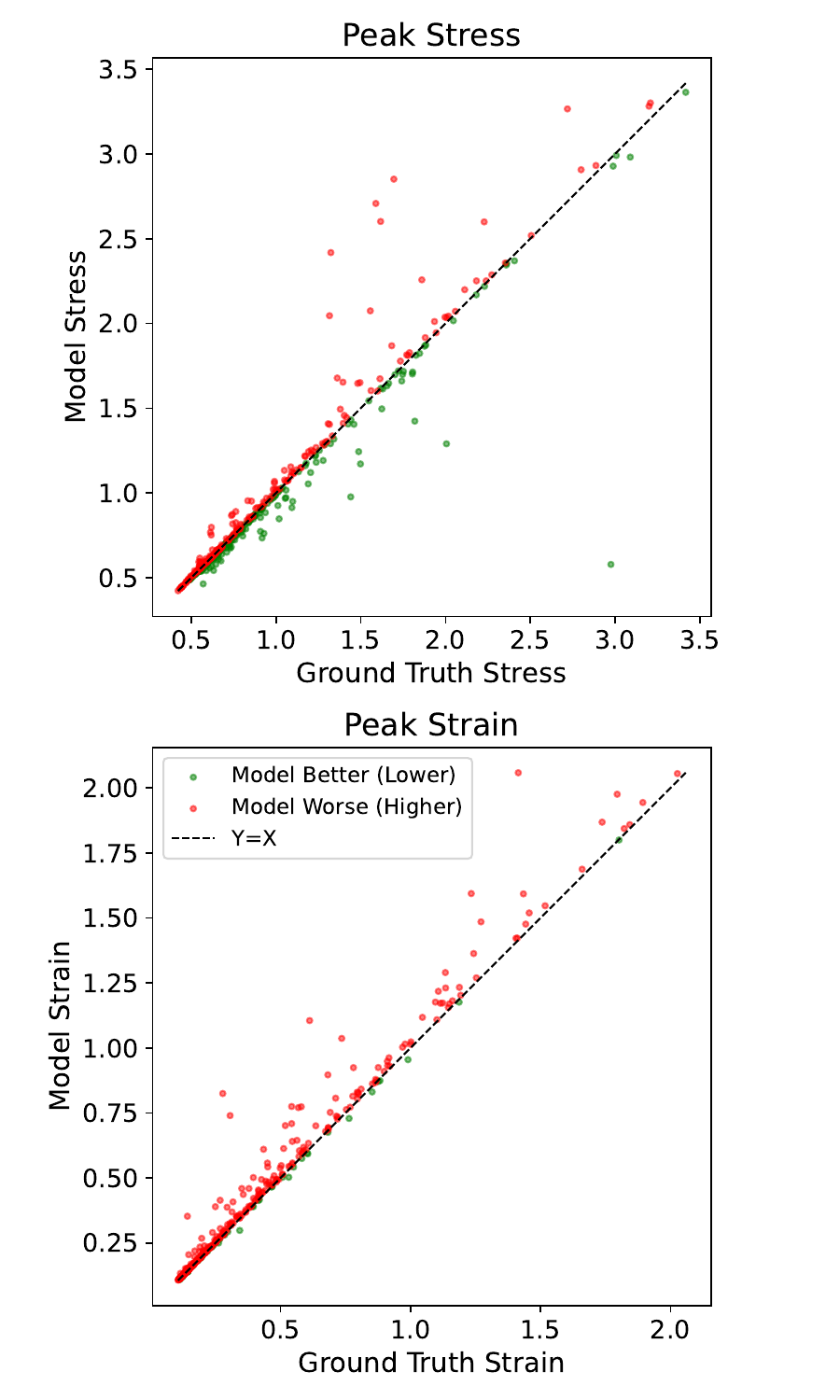}
    \captionof{figure}{Peak stress and strain scatter plots for model topology and ground truth topology. Each data point is a sample where red indicates the model's stress or strain is worse (higher) than the ground truth. Green indicates the model's stress or strain is better (lower) than the ground truth. Samples are from the validation dataset on the ViT-Small-4 model}
    \label{fig:stressstrain_stats}
\vspace{0.25cm}
\captionof{table}{Statistics for topology stress and strain, calculated on the model and ground truth topology for the ViT-Small-4 model.}

\resizebox{0.8\linewidth}{!}{%
\begin{tabular}{|l|r@{.}l|r@{.}l|}
\hline
\rowcolor[HTML]{C0C0C0} \textbf{Metric} & \multicolumn{2}{c|}{\textbf{Stress}} & \multicolumn{2}{c|}{\textbf{Strain}}\\ 
\hline
        Mean Absolute Error & 4&71 \% & 4&39 \% \\
        Std. Absolute Error & 8&43 \% & 7&49 \% \\
        Correlation & 0&979 & 0&989 \\
        Best Fit Slope & 1&032 & 1&189\\
        \hline

\end{tabular}
\label{table:ss_stats}%
}
\vspace{0.5cm}
\end{minipage}

\noindent fied. By backpropagating the gradients of this loss with respect to the predicted density field, i.e., $\frac{\partial FM(\tilde{\rho})}{\partial \tilde{\rho}_e}$, the topology can be adjusted to suppress floating material without requiring additional training. This approach provides a simple, fully differentiable mechanism for improving geometric consistency.

We apply this minimal post-processing step to the best-performing model, ViT-Small-4, to assess its effectiveness. The results are shown in \cref{fig:post-processing} and \cref{table:post-processing}. The floating material error is reduced significantly from 6.60\% to 0.8\%, indicating a substantial improvement in structural connectivity.
This reduction in disconnected regions also leads to a slight improvement in average compliance error, decreasing from 1.86\% to 1.75\%. However, the median compliance error and the proportion of high-error samples (above 30\%) remain unchanged. This suggests that while the post-processing step effectively removes superficial geometric artifacts, it does not correct fundamentally suboptimal structural layouts. In other words, the primary sources of error are associated with global design decisions rather than local connectivity issues.

The volume fraction error increases marginally by 0.17\%, which is negligible in practice. Overall, these results demonstrate that the proposed post-processing approach improves geometric quality with minimal impact on structural performance, while preserving the efficiency of the non-iterative framework.

\subsection{Generalization Analysis}\label{general_res}
\begin{figure}
        \centering
    \includegraphics[width=1\linewidth]{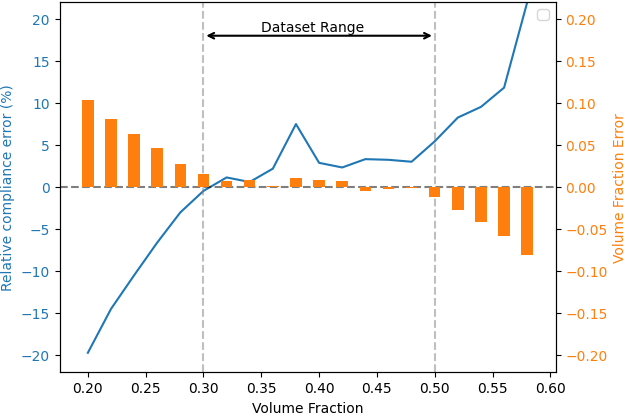}
    \caption{Compliance and volume fraction error results for the static ViT Small model. All samples are outside of the training dataset, with a varying input volume fraction. Range of values for the dataset are denoted on the plot.}
    \label{fig:generalization}
\end{figure}

To evaluate the model’s generalization capability, we analyze its response to variations in the prescribed volume fraction. Starting from a sample in the validation set, the input volume fraction is varied continuously over the range $0.2$ to $0.6$, while all other inputs are held fixed. The training dataset spans volume fractions between $0.3$ and $0.5$, so predictions within this interval correspond to interpolation, while values outside this range represent extrapolation. The results are shown in \cref{fig:generalization}.

Within the training range ($0.3$–$0.5$), the model exhibits strong performance, with volume fraction errors remaining within 1–3\% and compliance errors generally below 8\%. This indicates that the model accurately tracks changes in prescribed material usage and produces structurally consistent designs when operating within the distribution it has learned.

Outside the training range, however, systematic deviations emerge. For volume fractions below $0.3$, the model overestimates the amount of material, effectively saturating near the lower bound of the training distribution. This results in structures that are artificially stiff, leading to negative compliance error relative to the ground truth. Conversely, for volume fractions above 

\noindent\begin{minipage}{\linewidth}
\centering
    \includegraphics[width=\linewidth, trim={32mm 32mm 25mm 30mm}, clip]{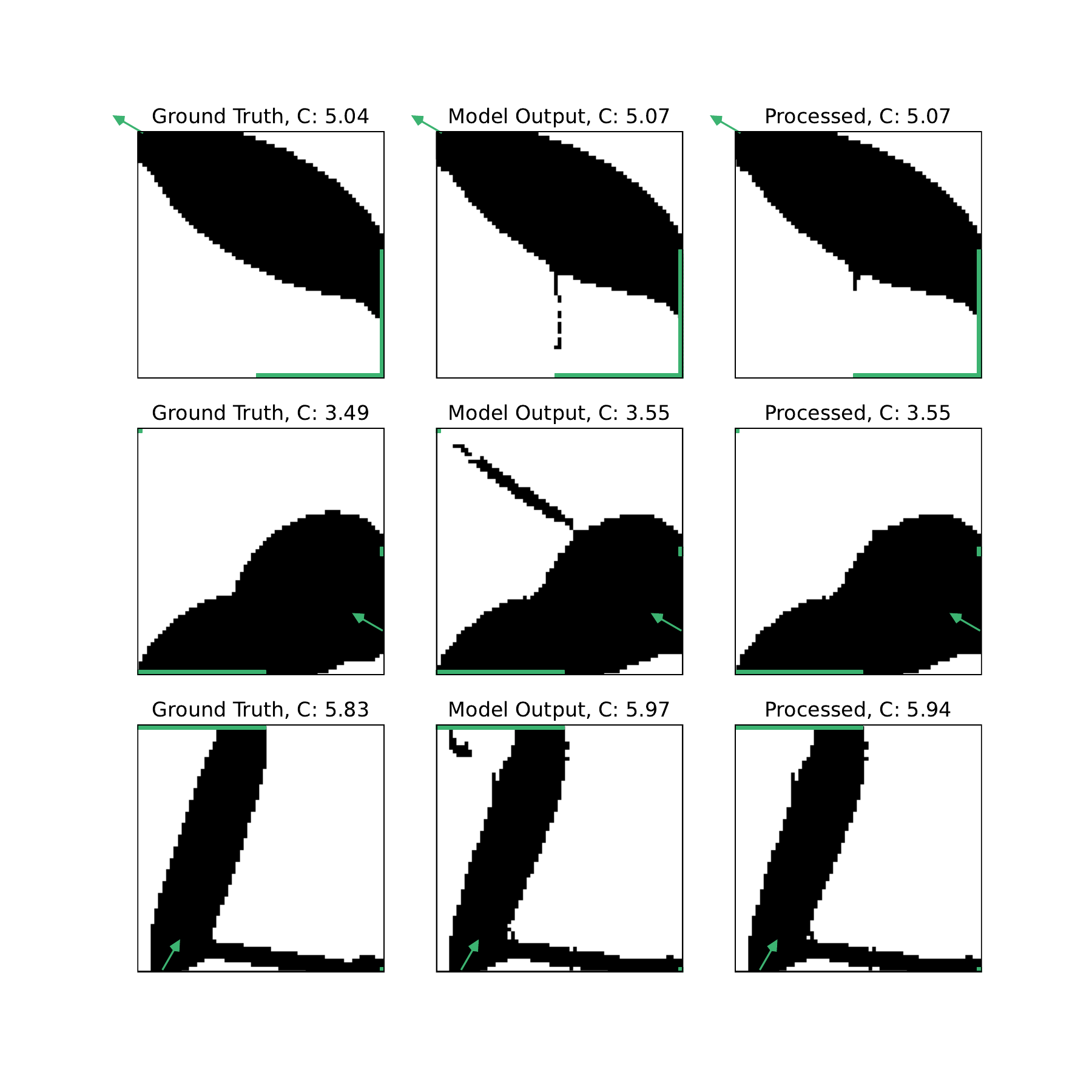}
    \captionof{figure}{Select samples from the post-processing results for the ViT-Small-4, showing how pixels are removed to improve floating material without consideration for compliance. Each topology is shown with its compliance C.}
    \label{fig:post-processing}
\vspace{0.25cm}
\captionof{table}{Post-processing results for the ViT-Small-4 model across the tracked metrics. Samples are first generated from the transformer model, then the floating material loss is used to calculate gradients to update the topology.}

\begin{tabular}{|l|l|l|}
\hline
\rowcolor[HTML]{C0C0C0} \textbf{Metric} & \textbf{ViT-S-4} & \textbf{Processed}\\ 
\hline
        Comp. Error (\%) & 1.86 & 1.75 \\
        Comp. Error $\geq$30\% (\%) & 2.20 & 2.20 \\
        Median Comp. Error (\%) & 0.32 & 0.32 \\
        Volume Fraction Error (\%) & 1.27 & 1.44 \\
        Load Discrepancy (\%) & 0.00 & 0.00 \\
        Floating Material Error (\%) & 6.60 & 0.80 \\
        \hline
\end{tabular}
\label{table:post-processing}%
\vspace{0.5cm}
\end{minipage} 

\noindent$0.5$, the model underestimates material usage, producing underbuilt structures with increasing compliance error. These trends are clearly reflected in \cref{fig:generalization}, where both compliance and volume fraction errors grow with distance from the training range.

These results indicate that the model does not extrapolate reliably beyond the training distribution and instead exhibits a strong inductive bias toward the range of volume fractions seen during training. While interpolation performance is robust, extrapolation leads to systematic errors in both material allocation and structural response.

\begin{figure}
        \centering
    \includegraphics[width=1\linewidth, trim={35mm, 15mm, 30mm, 0mm}, clip]{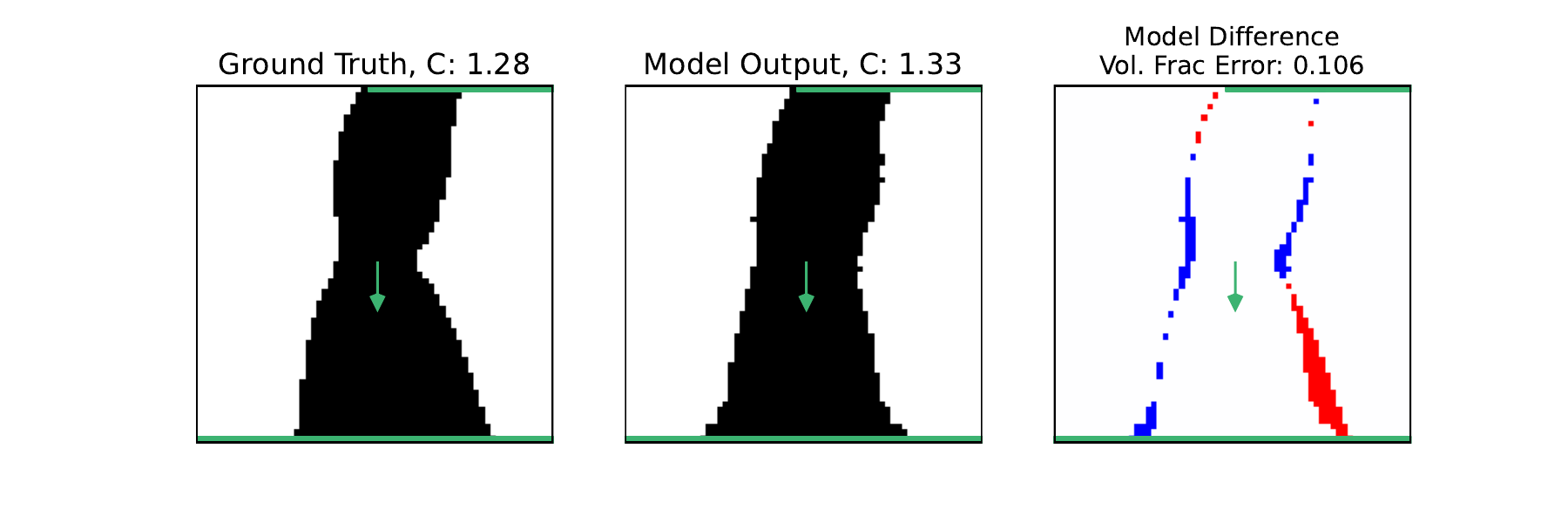}
    \caption{Example topology with an out of distribution load condition, a load in the center of the domain, sampled on the ViT-Small-4 model. The dataset only has loads placed on the outer perimeter of the domain.}
    \label{fig:midload}
\end{figure}

While the training dataset includes loads applied along the boundary of the domain, we also consider an out-of-distribution case with an internal load. As shown in \cref{fig:midload}, the ViT-Small-4 model generalizes well to this scenario, correctly placing material along load paths and maintaining close agreement with the prescribed volume fraction. This suggests that the model generalizes more effectively to unseen spatial load configurations than to unseen global constraints such as volume fraction.

\subsection{Dynamic Model Results}\label{dynamic_res}

\begin{table*}
\label{table:dynamic}
\begin{tabular}{|l|l|l|l|l|}
\hline
\rowcolor[HTML]{C0C0C0} 
\textbf{Metric} & \textbf{Untrained} & \textbf{Cond. Projection} & \textbf{Decoder Projection} & \textbf{Decoder Layers} \\
\hline
Compliance Error (\%)              & 15.14 & 12.72 & 12.93 & 4.81 \\
Compliance Error Above   30\% (\%) & 21.8  & 44.4  & 35.6  & 16.6 \\
Median Compliance Error   (\%)     & 4.18  & 4.00& 3.74  & 0.22 \\
Volume Fraction Error (\%)         & 1.62  & 6.39  & 7.49  & 4.38 \\
Load Discrepancy (\%)              & 5.00& 36.6  & 22.2  & 9.20\\
Floating Material (\%)             & 21.4  & 42.6  & 28.2  & 48.0\\ \hline \end{tabular}%
\caption{Analysis metrics for the dynamic topology optimization dataset on the transfer learning model, using ViT-Small as the pre-trained model. The untrained model is compared against the conditioning projection tuning, the decoder projection tuning, and the decoder layers tuning models.}
\end{table*}

We evaluate the proposed transformer framework on dynamic topology optimization problems. We adopt a transfer learning strategy, using the ViT-Small model trained on the static dataset as a baseline.
To incorporate temporal information, the conditioning token is augmented with frequency-domain features derived from the applied load via Fourier transformation as described in \sref{sec:4d}. We investigate three transfer learning strategies with increasing levels of model adaptation: (i) conditioning projection tuning, where only the input embedding is updated; (ii) decoder projection tuning, where the output mapping is refined; and (iii) decoder layer tuning, where both projection layers and the final transformer layers are fine-tuned. These configurations allow us to assess the trade-off between model flexibility and data efficiency.

Among these approaches, the decoder layer tuning strategy achieves the best overall performance. This model attains an average compliance error of 4.81\%, demonstrating that the transformer is capable of capturing the dominant structural response under dynamic loading conditions. Notably, this level of accuracy is achieved without iterative optimization, highlighting the effectiveness of the proposed operator-learning framework in extending to time-dependent problems.

\begin{figure}
        \centering
    \includegraphics[width=1\linewidth, trim={0mm, 9mm, 0mm, 2mm}, clip]{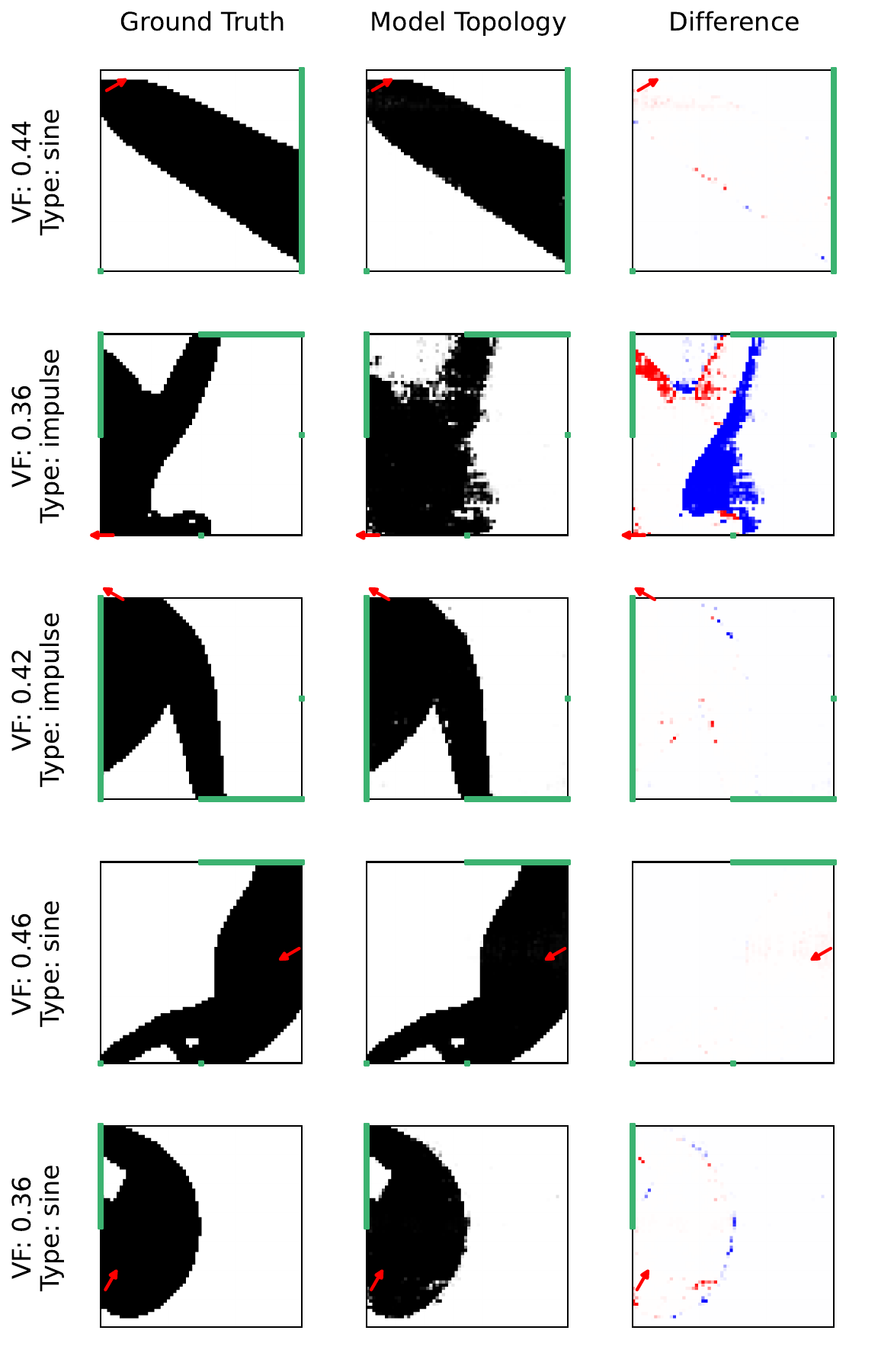}
    \caption{Validation samples from the ViT Small model finetuned on the decoder layers for dynamic data. The left column shows the ground truth topology, the middle column shows the model's output, and the right column shows the difference between the two.} 
    \label{fig:dyn_samples}
\end{figure}

However, the results also reveal a significant degradation in geometric quality compared to the static case. In particular, all dynamic models exhibit elevated floating material errors, with the best-performing model reaching approximately 48\%. This indicates that, while the predicted structures are mechanically functional in terms of compliance, they often contain disconnected or weakly connected regions. 
Samples of the generated topologies are shown in \cref{fig:dyn_samples}. 
Visual inspection of the generated topologies confirms that predictions exhibit blurred boundaries and reduced structural sharpness.

This discrepancy between mechanical performance and geometric fidelity suggests that the model successfully captures global load-response relationships but struggles to enforce strict topological constraints under limited data conditions. The dynamic setting introduces additional complexity through the temporal encoding, while simultaneously reducing the available training samples, which limits the model’s ability to learn robust connectivity patterns.

Despite these limitations, the proposed framework offers substantial computational advantages. The full inference pipeline, including model loading, initial FEA, and forward prediction, requires approximately 1.26 seconds, or 0.36 seconds when the model is preloaded. In contrast, classical SIMP-based dynamic topology optimization requires on the order of 8900 seconds. This corresponds to a speedup of over three orders of magnitude, enabling near real-time topology generation for dynamic problems.

These results demonstrate that transformer-based operator learning provides a viable path toward efficient dynamic topology optimization. While further improvements are needed to enhance geometric consistency and connectivity, the framework already achieves a compelling balance between accuracy and computational efficiency. Future work will focus on incorporating stronger inductive biases and constraint-aware training strategies to improve topology quality in dynamic settings.

\section{Conclusion}\label{sec:6}

In this work, we introduced a transformer-based operator learning framework for topology optimization that enables direct, non-iterative generation of structural designs from physics-informed inputs. By leveraging Vision Transformers to capture global spatial dependencies in stress and strain fields, and integrating auxiliary losses to enforce physical consistency, the proposed approach establishes a new paradigm for topology optimization that bypasses traditional iterative solvers.

The results demonstrate that transformer-based models can achieve competitive compliance accuracy while reducing inference time by multiple orders of magnitude compared to classical methods. In particular, the framework extends naturally to dynamic topology optimization through frequency-domain conditioning and transfer learning, enabling near real-time design generation even for time-dependent loading scenarios.
A key insight from this study is that topology optimization can be interpreted as a global operator learning problem, where optimal material distributions emerge from nonlocal interactions across the domain. Transformers provide a natural and effective architecture for capturing these interactions, offering a scalable alternative to both gradient-based solvers and iterative generative models.

Overall, this work highlights the potential of attention-based models to fundamentally transform topology optimization, shifting the field from iterative simulation-driven workflows toward fast, data-driven design generation.

\section{Limitations and Future Work}\label{sec:7}
While our highest performing transformer-based framework demonstrates high compliance accuracy for 2D structured domains, several extensions can be made to expand upon this work. An immediate improvement would be to adjust the architecture to allow for multi-load optimization. Our architecture only has parameters for encoding one load into a conditioning token, preventing the technique from working for multi-load scenarios. Extending the present approach to 3D topology optimization would greatly improve the applicability in real-world scenarios. Volumetric patchification (e.g. implementing  $8 \times 8 \times 8$ voxel patches) can be performed on 3D topology optimization samples, allowing the transformer architecture to be directly implemented. This would, however, increase the number of tokens dramatically, and increase the burden of the original projection layer. Future work would investigate other attention mechanisms that can more efficiently scale in 3D while preserving the long-range representational capacity that allowed this work to achieve accurate results.

 Dataset choices were made to maintain consistency with benchmark datasets, such as TopologyGAN and TopoDiff. This choice limits the generalization of the model to lower volume fractions or internal loads. Future work can explore adaptive or performance-based formulations rather than fixed volume ratios \cite{liang2002performance}.

Unstructured and adaptive meshes are widely used in practical topology optimization to capture geometric complexity and enable localized refinement in regions of high stress or design sensitivity. The current Vision Transformer formulation relies on a structured grid representation, which limits direct applicability to unstructured domains and restricts the ability to perform adaptive, resolution-varying refinement. While higher resolutions can be used within a regular grid, they do not provide the efficiency and flexibility of mesh adaptivity.
Graph-based neural networks offer a more natural representation for unstructured meshes and have shown promise in topology optimization \cite{tabarraei2025graph}. However, fully generative, physics-consistent models for unstructured topology design remain limited and represent an important direction for future research.

In addition, while the proposed framework achieves strong compliance accuracy, it exhibits challenges in enforcing global structural connectivity, particularly in dynamic settings. This highlights the need for improved inductive biases and constraint-aware learning strategies to enhance the physical consistency of generated designs. Overall, future work should focus on improving scalability, geometric fidelity, and robustness to enable broader applicability in real-world engineering design problems.

\subsection*{Declarations}
\noindent \textbf{Conflict of interest} On behalf of all authors, the corresponding author states that there is no conflict of interest.

\noindent \textbf{Funding} This work has been financially supported by the Institute of Digital Engineering - USA.

\noindent \textbf{Author contributions} Conceptualization: Srijan Das, Alireza Tabarraei; Methodology: Aaron Lutheran; Formal analysis and investigation: Aaron Lutheran; Writing - original draft preparation: Aaron Lutheran; Writing - review and editing: Srijan Das, Alireza Tabarraei; Supervision, Srijan Das, Alireza Tabarraei.

\noindent \textbf{Ethics approval and Consent to participate} Not applicable for this work.

\noindent \textbf{Data Availability} Dataset will be made available on request to the corresponding author.

\noindent \textbf{Replication of results} Replication material, including model parameters, and code, are available on request to the corresponding author.

\nocite{*}
\bibliographystyle{elsarticle-num}
\bibliography{Transformer-Model.bib}

\end{document}